\documentclass[9pt,twocolumn,a4paper]{article}
\usepackage[T1]{fontenc}
\usepackage{graphicx}
\usepackage[british,UKenglish,USenglish,english,american]{babel}

\begin{document}

\twocolumn[
  \begin{@twocolumnfalse}
    \begin{center}
    \huge \textbf{Bi-directional top hat D-Scan: single beam accurate characterization of nonlinear waveguides}\\ %
    \vspace{0.5cm}
    \large \textbf{Samuel Serna \textsuperscript{1,2}, Nicolas Dubreuil \textsuperscript{2,*}}\\ %
\normalsize \textit{\textsuperscript{1}Centre for Nanoscience and Nanotechnology - C2N site-Orsay, CNRS UMR 8622, Universit\'{e} Paris Saclay, Bat. 220, 91405 Orsay Cedex, France\\ %
\textsuperscript{2}Laboratoire Charles Fabry, Institut d'Optique Graduate School,
CNRS, Universit\'{e} Paris-Saclay, 91127 Palaiseau cedex, France \\ %
\textsuperscript{*}Corresponding author: nicolas.dubreuil@institutoptique.fr
}
\end{center}

    \begin{abstract}
      \textbf{The characterization of a third order nonlinear integrated waveguide is reported for the first time by means of a top-hat Dispersive-Scan (D-Scan) technique, a temporal analog of the top-hat Z-Scan. With a single laser beam, and by carrying two counter-directional nonlinear transmissions to assess the input and output coupling efficiencies, a novel procedure is described leading to an accurate measurement of the TPA figure of merit, the effective Two-Photon Absorption (TPA) and optical Kerr (including the sign) coefficients. The technique is validated in a silicon strip waveguide for which the effective nonlinear coefficients are measured with an accuracy of $\pm 10~\%$.}    \end{abstract}
  \end{@twocolumnfalse}
]



Recent progresses in silicon photonics technologies have enabled the realization of passive and active optical components for applications in optical fibre communications, optical interconnects inside and between CMOS chips, and in biophotonics \cite{VivienPavesi:2013fk}. One of the major future challenges concern the development of compact optical devices for all-optical signal processing based on nonlinear interactions. While the optical nonlinearities in silicon based waveguides or micro-cavities have been extensively studied \cite{Tsang:08,Osgood:09}, their performances at telecom wavelengths suffer from an excess of nonlinear losses imposed by the two-photon absorption (TPA) and the free-carrier absorption (FCA) processes \cite{yin:07}. To overcome this limitation, integrated structures have been fabricated in materials with higher energy bandgap, such as AlGaAs \cite{Oda:07}, GaInP \cite{Combrie:09} or chalcogenide glasses \cite{Suzuki:09}. In order to preserve the compatibility with CMOS processes, alternative materials like amorphous silicon \cite{Kuyken:11} 
or SiN \cite{Ikeda:08,lacava2017si-rich} are investigated to achieve low-TPA nonlinear waveguide structures. 
Finally, one can cite hollow core silicon waveguides 
that can be filled with nonlinear polymer based material \cite{Vallaitis:09}. 

Nonlinear response in waveguides is evaluated through an effective nonlinear parameter $\gamma$, where the real part $Re(\gamma)$ refers to optical Kerr effect and the imaginary part $Im(\gamma)$ to TPA. Various techniques based on Self-Phase Modulation (SPM), Cross-Phase Modulation or Four-Wave-Mixing effects \cite{Agrawal2006424}, are used to assess the effective Kerr parameter $Re(\gamma)$ through the nonlinear phase shift measurement $\phi_{NL0}=Re(\gamma) P_{0}L_{eff}$, with $P_{0}$ the injected optical power, $L_{eff}$ the waveguide effective length. Similarly, the effective TPA parameter  $Im(\gamma)$ is evaluated by measuring the transmission of pulses as a function of the input power. Beyond the techniques applied to characterize nonlinear waveguides, the determination of the parameter $\gamma$ from experimental data requires a prior knowledge of the optical peak power $P_{0}$ injected inside the waveguide. Whereas, the coupling efficiency determination in optical fibres can be easily achieved, for instance through a cut-back technique, it turns to be a real difficulty with sub-micron size waveguides. Nonetheless, third order nonlinear characterizations of integrated waveguides currently reported in the literature rarely give a precise determination of the coupling efficiency which is essential to quantify accurately the nonlinear waveguide properties. 

Using a single beam top-hat Dispersive-Scan (D-Scan) method, we present a first waveguide nonlinear characterization that simultaneously and accurately measures the input ($\kappa_{in}$) and output ($\kappa_{out}$) coupling efficiencies, the effective TPA and Kerr nonlinear parameters, $\gamma=Re(\gamma)+\imath Im(\gamma)$, including the sign of $Re(\gamma)$. Furthermore, our procedure enables to measure the nonlinear third order figure of merit $FOM_{\textrm{TPA}}=Re(\gamma)/(4\pi Im(\gamma))$ independently of the knowledge of $\kappa_{in}$, $\kappa_{out}$ and $L_{eff}$.

Due to its simplicity and performance, the Z-scan technique is currently the preferred method to characterize the third order nonlinear susceptibility of materials \cite{Sheik-bahae:89}, nevertheless, it can not be applied to waveguides. By keeping the advantage of using a single beam, this technique has been transposed into the temporal domain in order to characterize the nonlinear refractive indices of optical fibres \cite{Louradour:99, Fonseca:01, Lopez-Lago:01}. The D-Scan technique consists in measuring the SPM induced output spectral broadening of transmitted pulses for various dispersion coefficients $\phi^{(2)}$ applied to the injected pulses. Like in Z-Scan, D-Scan advantageously gives the sign of the Kerr coefficient and, with a single laser beam, a simultaneous access to the Kerr and TPA properties. 

\begin{figure}[htbp]
\centering
\includegraphics[width=\linewidth]{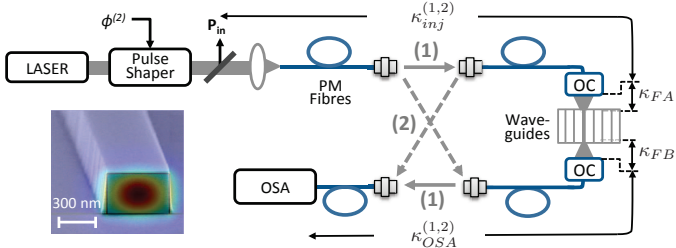}
\caption{Bi-directional top-hat D-Scan set-up for the nonlinear characterization of waveguides where dispersed pulses at 1580 nm are injected in a waveguide using an optical coupler (OC) connected to single-mode polarization-maintaining (PM) fibres. The transmitted pulses are analyzed with an optical spectrum analyzer (OSA). Inset: SEM waveguide image of a silicon strip waveguide with the simulated mode profile.}
\label{FigMontageDScan}
\end{figure}

To the best of our knowledge, D-Scan technique has never been used in a waveguide supporting TPA. In addition, we describe hereafter the first experimental demonstration of a top-hat D-Scan, that uses pulses with a quasi-rectangular spectral shape  to improve the sensitivity of the measurement, being the analogous in the temporal domain of the top-hat Z-scan~\cite{Zhao:93, Cherukulappurath:04}.

The experimental set-up, described in Fig.~\ref{FigMontageDScan}, uses a mode locked Erbium doped fibre laser that delivers 150 fs duration pulses, with a repetition rate of $F=50$~MHz. These pulses are sent through a grating-based pulse shaper that fixes the pulse spectrum following a quasi-rectangular shape of 7.3 nm width nearby $\lambda=1585$~nm and introduces an adjustable dispersion coefficient $\phi^{(2)}$ comprised between -3 to +3~ps$^2$. The design and characterization of the pulse shaper is detailed in \cite{Serna:15}. For $\phi^{(2)}=0$ ps$^2$, the autocorrelation pulse duration is measured equals to $T_0=2$~ps, close to the Fourier limit. By means of single mode Polarization-Maintaining (PM) fibres connected to an optical coupler (OC), pulses are injected inside a semiconductor waveguide. The optical coupler comprised a set of two-microscope objectives, between which a half-wave plate followed by a polarization beam splitter (PBS) cube are inserted to match with the polarization state of the fundamental TE waveguide mode.  Using a similar coupler, the transmitted pulses are injected in PM fibres connected to an Optical Spectrum Analyzer (OSA). The connections between the set of PM fibre patch cables allow achieving two counter-directional nonlinear transmissions, (1) and (2) as shown in Fig.~\ref{FigMontageDScan}.

In order to validate our experiment, we choose to characterize a $L=1$~cm long silicon-on-insulator strip waveguide, material that has been thoroughly studied in the nonlinear regime at telecom wavelengths \cite{Tsang:08}. A SEM picture of the waveguide, with a 340$\times$520~nm$^2$ section, overlapped with the simulation mode profile is shown in Fig.~\ref{FigMontageDScan} (Inset). By analyzing the spectral Fabry-Perot interference fringes on the spectral transmission achieved with a tunable laser diode, the linear propagation losses of the waveguide are measured equal to $\alpha=1.0\pm 0.1$~cm$^{-1}$. The variation with wavelength of the spectral interval between the fringes enables to measure the mode group index $n_g=4.2$.

\begin{figure}[htbp]
\centering
\includegraphics[width=\linewidth]{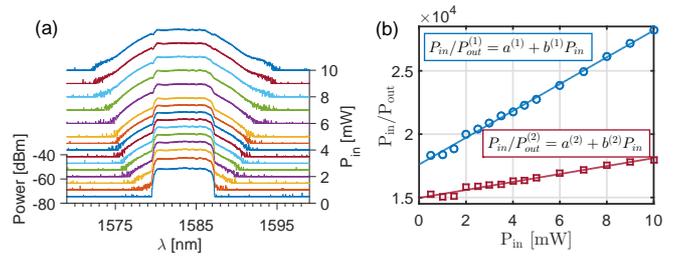}
\caption{For incident power varying from 0.5 to 10~mW and setting $\phi^{(2)}=0$ ps$^2$, (a) measured output spectra  and (b) $P_{\textrm{in}}/P_{\textrm{out}}$ curves as a function of $P_{\textrm{in}}$ measured for the two injection directions (1) and (2). }
\label{FigOutSpectra_PinPout}
\end{figure}

The PM fibre patch cables are first connected in order to direct the injection from the waveguide facets A to B, direction (1). By fixing $\phi^{(2)}=0$ ps$^2$ and for incident average powers $P_{\textrm{in}}$ varying from 0.5 to 10~mW, the output spectra are recorded and depicted in Fig.~\ref{FigOutSpectra_PinPout}(a). As the input power increases, output spectra exhibit a symmetric spectral broadening, characteristic from SPM induced by optical Kerr effect. Under such a pulse regime, no blue shift is observed on the  spectra showing that the free-carier refraction effect can be neglected in our experiment. From the recorded output spectra, the transmitted output average power $P_{\textrm{out}}^{(1)}$ is calculated and the ratio $P_{\textrm{in}}/P_{\textrm{out}}^{(1)}$ is plotted with open blue circles in Fig.~\ref{FigOutSpectra_PinPout}(b) as a function of $P_{\textrm{in}}$. As expected from TPA effect, the measured points follow a linear evolution $P_{\textrm{in}}/P_{\textrm{out}}^{(1)}=a^{(1)}+b^{(1)}P_{\textrm{in}}$, where the linear fit gives $a^{(1)} = 17605 \pm 1\%$, $b^{(1)} = 1059$~mW$^{-1}$ $\pm 3\%$. These coefficients are actually related to the input and output coupling efficiencies, $\kappa_{\textrm{in}}^{(1)}$ and $\kappa_{\textrm{out}}^{(1)}$, the linear propagation losses $\alpha$ and the effective length $L_{eff}=(1-\exp(-\alpha L))/\alpha$, through the relations \cite{Baron:09}: $a^{(1)}=1/(\kappa_{\textrm{in}}^{(1)}\kappa_{\textrm{out}}^{(1)}\exp(-\alpha L))$, $b^{(1)}=2 Im(\gamma) L_{\textrm{eff}}\eta/(\kappa_{\textrm{out}}^{(1)}\exp(-\alpha L))$, where $\eta=1/(F \int^{1/F}_0\left| U(t)\right| ^2dt$) and $U(t)$  the normalized temporal shape of the pulse. 

One underlines that the fit values $a^{(1)}$ and $b^{(1)}$ can not assess the TPA parameter without measuring the input coupling efficiency as $Im(\gamma)= b^{(1)}/(2 a^{(1)}\kappa_{\textrm{in}}^{(1)}\eta L_{eff})$.  
Actually, the input and output coupling efficiencies account for the coupling coefficients at the facet A and B, $\kappa_{\textrm{FA}}$ and $\kappa_{\textrm{FB}}$, and the input and output transmissions $\kappa_{\textrm{inj}}^{(1)}$ and $\kappa_{\textrm{OSA}}^{(1)}$ (see Fig.~\ref{FigMontageDScan}), through the relations  $\kappa_{\textrm{in}}^{(1)}=\kappa_{\textrm{inj}}^{(1)}\kappa_{\textrm{FA}}$ and $\kappa_{\textrm{out}}^{(1)}=\kappa_{\textrm{FB}}\kappa_{\textrm{OSA}}^{(1)}$. The losses $\kappa_{\textrm{inj}}^{(1)}$ and $\kappa_{\textrm{OSA}}^{(1)}$ are measured on the bench. 
In order to determine $\kappa_{\textrm{FA}}$ and $\kappa_{\textrm{FB}}$, and keeping the injection alignement in place, the laser injection is reversed along the direction (2) (see Fig.~\ref{FigMontageDScan}). For $\phi^{(2)}=0~\mathrm{ps}^2$, the measured transmission curve $P_{\textrm{in}}/P_{\textrm{out}}^{(2)}$ as a function of $P_{\textrm{in}}$ is depicted in Fig.~\ref{FigOutSpectra_PinPout}(b) with open red squares and the related linear fit coefficients $a^{(2)}$ and $b^{(2)}$ are evaluated respectively equal to $14967\pm 1\%$ and $314$~mW$^{-1}$ $\pm 10\%$.  Similarly to the direction (1), the fit coefficients depend on input and output coupling efficiencies $\kappa_{\textrm{in}}^{(2)}=\kappa_{\textrm{inj}}^{(2)}\kappa_{\textrm{FB}}$ and $\kappa_{\textrm{out}}^{(2)}=\kappa_{\textrm{FA}}\kappa_{\textrm{OSA}}^{(2)}$, for which $\kappa_{\textrm{inj}}^{(2)}$ and $\kappa_{\textrm{OSA}}^{(2)}$ are measured on the bench.
Assuming that the coupling coefficients from each facet are independent of the direction of propagation, the coupling efficiencies can be deduced from the equalities $(\kappa_{\textrm{FA}})^2=b^{(1)}/(b^{(2)}a^{(1)}\kappa_{\textrm{inj}}^{(1)}\kappa_{\textrm{OSA}}^{(2)}\exp(-\alpha L))$ and $(\kappa_{\textrm{FB}})^2=b^{(2)}/(b^{(1)}a^{(2)}\kappa_{\textrm{inj}}^{(2)}\kappa_{\textrm{OSA}}^{(1)}\exp(-\alpha L))$, which only depend on measured parameters. The values are found equal to $\kappa_{\textrm{FA}}=6.4~\%$ and $\kappa_{\textrm{FB}}=2.5~\%$, with an  uncertainty of $\pm~5~\%$ calculated from the uncertainties provided for each measured coefficient. Knowing the input coupling coefficient, the TPA parameter is then evaluated through $Im(\gamma)= b^{(1)}/(2 a^{(1)}\kappa_{\textrm{in}}^{(1)}\eta L_{eff})$ and found equal to $(16\pm2)~\textrm{W}^{-1}\textrm{m}^{-1}$.

\begin{figure}[b]
\centering
\includegraphics[width=\linewidth]{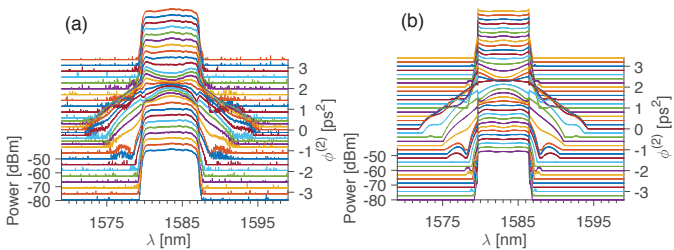}
\caption{(a) Experimental spectra registered at $P_{in}=10$mW and varying the second order dispersion between -3 to +3~$\textrm{ps}^2$. (b) Output spectra calculated from the relation (\ref{Eq_NLphase}) with the parameters assessed throughout the experience.}
\label{FigSimuExp}
\end{figure}

Our objective is now to measure the figure of merit of the waveguide to assess it effective Kerr coefficient.
Both Z-Scan and D-Scan techniques consist in analyzing the intensity dependent nonlinear phase shift and absorption of materials by varying the beam size, respectively in the spatial and temporal domains. Whereas the spatial beam size is governed by diffraction effect in Z-Scan, the pulse duration is varied through an adjustable 2nd order dispersion coefficient in D-Scan. 
The spectral output measurements are then repeated for dispersion coefficient $\phi^{(2)}$ set between -3 to +3~$\textrm{ps}^2$. The cumulative output spectra recorded for $P_{in}=10$ mW are plotted in Fig.~\ref{FigSimuExp}(a) with a vertical log-scale extending a dynamic of 30~dB. The largest spectral broadenings are achieved nearby  $\phi^{(2)}=0~\textrm{ps}^2$ as the pulse duration is minimized, while the peak power reaches its maximum value. For larger dispersion, the pulse duration exceeds 12~ps and the spectral broadening vanishes. 

\begin{figure}[t]
\centering
\includegraphics[width=8.4cm]{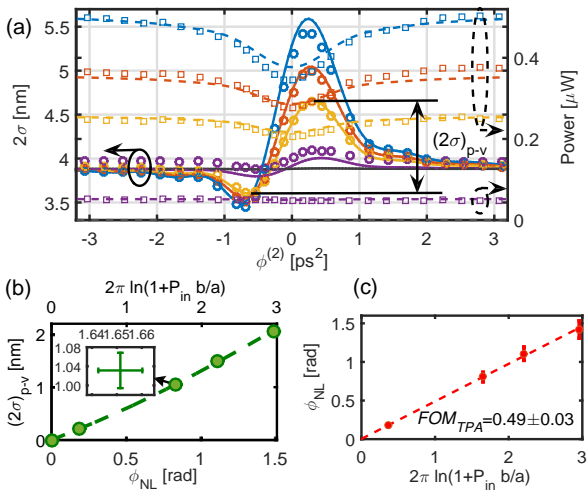}
\caption{(a) Spectral r.m.s. linewidth $2\sigma$ (open circles) and output power $P_{\textrm{out}}$ (open squares) measurements with the dispersion coefficient $\phi^{(2)}$ applied on the incident pulse for $P_{\textrm{in}}=1, 5, 7$ and 10~mW. Their calculated variations are plotted respectively with solid and dashed lines. (b) Experimental (dots) and simulated (dashed line) peak to valley values as a function of the measured parameter $2\pi \ln(1+P_{\textrm{in}}b/a)$ and the nonlinear phase shift $\phi_{NL}$ respectively. Inset: measured point with related error bars. (c) Measurement of the figure of merit $FOM_{\textrm{TPA}}$  using a linear fit on the variation between $\phi_{NL}$ and $2\pi \ln\left(1+P_{in} b/a\right)$.}
\label{FigDScanFOM}
\end{figure}

For a closer analysis, the evolutions of the output power $P_{\textrm{out}}^{(1)}$ (open squares) and the spectral r.m.s.~linewidth $2\sigma$ (open circles), which are extracted from the transmitted spectra, are presented in Fig.~\ref{FigDScanFOM}(a) as a function of $\phi^{(2)}$ and for  input average powers $P_{\textrm{in}}=1, 5, 7$ and 10~mW. The $2\sigma$ curves present a dispersive shape and the output power curves symmetrically decrease towards $\phi^{(2)}=0~\textrm{ps}^2$, following the behaviors reported in Z-Scan transmission curves \cite{Sheik-bahae:89}. 
For large dispersion coefficients, either positive or negative, the pulse duration is too large to efficiently generate SPM induced spectral broadening and TPA effects. The $2\sigma$ curves asymptotically tend to the spectral linewidth of the incident pulse, while $P_{\textrm{out}}$ curves reach the linear regime values. For smaller dispersion coefficients, $P_{\textrm{out}}$ curves present a symmetric dip centered at $\phi^{(2)}=0~\textrm{ps}^2$, signifying that absorption raises with an increase of the pulse peak intensity in accordance with TPA. Simultaneously, the spectral broadening is larger as the dispersion is closer to 0. In agreement with the positive sign of $n_2$ for silicon, the maximum spectral broadening is observed towards positive $\phi^{(2)}$ values. The temporal Kerr lens effect increments the curvature of the temporal wave front of the chirped pulse. The spectrum is broader and reaches a maximum nearby  $\phi^{(2)}\simeq+0.25~\textrm{ps}^2$. On the other hand, for negative $\phi^{(2)}$, the Kerr lens counter-balances the dispersion effect by inducing a temporal wave front opposite in sign to that of the chirped pulse, which modifies the spectrum giving rise to a reduced linewidth.  Similarly to Z-Scan, the peak-to-valley deviation $(2\sigma)_{\textrm{P-V}}$ on each $2\sigma$ curves, identified in Fig.~\ref{FigDScanFOM}(a), is directly related to the strength of SPM and can be used to retrieve the nonlinear phase shift experienced by the pulse \cite{Louradour:99}. 
 
To perform such analysis, we have simulated the propagation of chirped pulses through a nonlinear waveguide supporting both optical Kerr and TPA effects.  Considering the geometry of the SOI strip waveguide with a $0.12~\mu$m$^2$ effective area, one can neglect the dispersion effect in a 1~cm long waveguide and for 2~ps pulse duration as the dispersion length is estimated equal to 0.5~m. In accordance with the rectangular like shape of the spectrum, the pulse envelop shape is set to $\mathrm{Sinc}(a t)$, where $a=2.317~\mathrm{ps}^{-1}$ for consistence with the autocorrelation measurement. The dispersed pulse shape is calculated by applying, in the Fourier domain, the transfert function $\exp( i\phi^{(2)}\omega^{2}/2)$ for a second order dispersive linear medium. Following an inverse Fourier transform, the pulse envelop $A(0,t)=I_0 U(t)$ injected inside the waveguide is determined, with $I_0$ the injected intensity. Neglecting the free-carrier effects, justified by the symmetric spectral broadening, the pulse envelope at the output waveguide is given by $A(L,t)=\sqrt{I(L,t)}\exp(i\phi_{NL}(L,t))$, where the intensity and  nonlinear phase shift, driven by TPA and optical Kerr effects,  follow the relations  \cite{yin:07}:
\begin{equation}\label{Eq_intensity}
I(L,t)=\frac{I(0,t)\exp(-\alpha L)}{1+\beta_{TPA}I(0,t)L_{\mathrm{eff}}},
\end{equation} 
\begin{equation}\label{Eq_NLphase}
\phi_{NL}(L,t)=k_0\frac{n_{2}}{\beta_{TPA}}\ln \left(1+\beta_{TPA}I(0,t)L_{\mathrm{eff}} \right),
\end{equation}
where $I(0,t)=|A(0,t)|^2$. The nonlinear refractive index $n_2$ and the TPA coefficient $\beta_{TPA}$ are related respectively to $Re(\gamma)=k_0 S^2n_2/ A_{\textrm{NL}}$ and $Im(\gamma)= S^2\beta_{\textrm{TPA}}/ (2 A_{\textrm{NL}})$, with $A_{NL}$ the nonlinear effective area of the waveguide and $k_0=\frac{2\pi}{\lambda}$. Note that both relations include the slow down factor  given by the ratio $S= n_{\textrm{g}}/n_{\textrm{0}}$, between the group index of the mode and the refractive index of the waveguide material to account for a rigorous definition of the effective nonlinear parameters \cite{Baron:09, Sato:15}. 
Using the relations (\ref{Eq_intensity}) and (\ref{Eq_NLphase}), the pulse output spectra are computed and the spectral r.m.s. linewidth $2\sigma$ is calculated for various dispersion coefficients $\phi^{(2)}$ and nonlinear phase shifts $\phi_{NL}=\phi_{NL}(L,0)$.
The evolution of the calculated peak-to-valley difference $(2\sigma)_{\textrm{P-V}}$ 
with the accumulated nonlinear phase shift $\phi_{\textrm{NL}}$ is plotted in Fig.~\ref{FigDScanFOM}~(b) with dashed lines, which shows a quasi-linear relation. 

To determine the experimental $(2\sigma)_{\textrm{P-V}}$ value for each power, a parabolic fit is applied on a subset of points nearby the maximum and minimum of the $2\sigma$ curve.  The error bars, shown for one point (see inset), are deduced from the 95~\% confidence intervals for the fit coefficients.
Knowing that the nonlinear phase shift under TPA follows $\phi_{\textrm{NL}}=2\pi FOM_{\textrm{TPA}}\ln\left(1+\beta_{\textrm{TPA}}I_0 L_{\textrm{eff}}\right)$ and that $\beta_{\textrm{TPA}}I_0 L_{\textrm{eff}}=P_{\textrm{in}}b^{(1)}/a^{(1)}$ (in accordance with (\ref{Eq_NLphase})), the measured peak-to-valley difference $(2\sigma)_{\textrm{P-V}}$ are reported on the graph in Fig.~\ref{FigDScanFOM}~(b) with dots as a function of $2\pi \ln(1+P_{\textrm{in}}b^{(1)}/a^{(1)})$. 
By combining the calculated $(2\sigma)_{\textrm{P-V}}$ variation with $\phi_{\textrm{NL}}$ and the  $(2\sigma)_{\textrm{P-V}}$ values measured for different incident power, the nonlinear phase shifts $\phi_{\textrm{NL}}$ achieved experimentally are reported in Fig.~\ref{FigDScanFOM}~(c) with filled dots in terms of the experimental parameter $2\pi \ln(1+P_{\textrm{in}}b^{(1)}/a^{(1)})$. Whereas the error bars on the horizontal axis are negligible, the measured $\phi_{\textrm{NL}}$ values are given with uncertainties of $\pm 10 \%$. As anticipated from the model, the nonlinear phase shift experienced by the pulse exhibits a linear variation with $2\pi \ln(1+P_{\textrm{in}}b^{(1)}/a^{(1)})$. The slope of the linear fit plotted with dashed line measures a $FOM_{\textrm{TPA}}$ for silicon equal to 0.49$\pm$0.03, which is in agreement with the values reported for silicon nearby $\lambda=1.5~\mu$m \cite{Tsang:08}.   

Now and using the relation $Re(\gamma)= 4\pi FOM_{\textrm{TPA}}Im(\gamma)$, the measured values for $FOM_{\textrm{TPA}}$ and $Im(\gamma)$, gives rise to the effective Kerr coefficient of the waveguide $Re(\gamma)=(+98\pm10)~\textrm{W}^{-1}\textrm{m}^{-1}$.
By simulating the modal field distribution in the waveguide, the effective nonlinear area is calculated equal to $A_{NL}=0.12~\mu\textrm{m}^2$. With a slow down factor $S=1.2$, our measurement procedure allows to determine $n_2 = +2.0 \times 10^{-18}$~m$^2$/W and $\beta_{\textrm{TPA}} = 2.6 \times 10^{-12}$~m/W, which are in very good agreement with the standard coefficients used in silicon \cite{Tsang:08}. 

To verify the consistency of the analysis with the measured curves, we have calculated the output spectra and powers by  incorporating the experimental parameters in the relations (\ref{Eq_intensity}) and (\ref{Eq_NLphase}). At $P_{in}=10$~mW, the superimposition of the calculated output spectra is plotted in Fig.~\ref{FigSimuExp}(b) for $\phi^{(2)}$ varying between -3 to +3~$\textrm{ps}^2$. The calculated spectra perfectly reproduce the features of the measured spectra shown in Fig.~\ref{FigSimuExp}(a), within the 30~dB experimental power scale. For input powers $P_{\textrm{in}}=1, 5, 7$ and 10~mW, the calculated r.m.s. spectral linewidths and output powers with $\phi^{(2)}$ are reported in  Fig.~\ref{FigDScanFOM}(a), respectively with  solid and dashed lines. The agreement with the measured values is very good, although slight discrepancies can be observed at 10~mW towards $\phi^{(2)}=0~\textrm{ps}^2$. We attribute these variations to the limit of our approximation concerning the pulse shape. Indeed, the relation between the measured average power $P_{\textrm{out}}$  and the output peak intensity given by (\ref{Eq_intensity}) is achieved through the parameter $\eta$ related to the incident pulse. A more accurate analysis would be necessary to account for the TPA induced temporal deformation, which is dependent on the input power but do not exceed $10~\%$ variations in our experience. 
However, the correspondance between the measured and calculated curves is achieved without any varying parameter.

In summary, we have reported an accurate measurement of the effective Kerr and Two-Photon Absorption coefficients of an integrated waveguide by means of a top-hat Dispersive-Scan (D-Scan) method, a temporal analog of the top-hat Z-scan. Our procedure allows, with one single laser beam, to experimentally assess the nonlinear third order figure of merit $FOM_{\textrm{TPA}}$ of the waveguide independently of the input and output coupling efficiencies, and of its effective length. Furthermore, by carrying two counter-directional nonlinear transmissions,  the coupling efficiencies at the two facets of the waveguide are determined, enabling the measurement of  the effective TPA and Kerr coefficients with an accuracy of $\pm 10~\%$. The technique has been validated in a silicon strip waveguide, for which we retrieve the nonlinear coefficients given in the literature for silicon. By experimentally assessing the coupling efficiencies, a difficult issue in integrated optics, our method contributes in improving the methodologies to accurately characterize and compare the nonlinear performances  between integrated nonlinear devices. 

\section*{Acknowledgment.} The authors thank Julia Danchenko, Laurent Vivien, Eric Cassan and Philippe Delaye for helpful discussions about the experimental data analysis.

\bibliography{BibTopHatDScan}

\end{document}